# White Paper: The Generative Education (GenEd) Framework

Empowering Educators as AI-Enhanced Mentors

Author:
Daniel Leiker
Founder, GenEd | [www.gened.ai](www.gened.ai)
PhD Researcher, UCL Knowledge Lab
Chief E-Learning Officer, Immerse Education

## Abstract

The Generative Education (GenEd) Framework explores the transition from Large Language Models (LLMs) to Large Multimodal Models (LMMs) in education, envisioning a harmonious relationship between AI and educators to enhance learning experiences. This paper delves into the potential of LMMs to create personalized, interactive, and emotionally-aware learning environments. Through addressing the Two-Sigma problem and the introduction of a conceptual product named Harmony, the narrative emphasizes educator development, adapting policy frameworks, and fostering cross-sector collaboration to realize the envisioned AI-enhanced education landscape. The discussion underscores the urgency for proactive adaptation amidst AI's evolution, offering a pragmatic roadmap to navigate the technical, ethical, and policy intricacies of integrating AI in education.

# Table of Contents

# Introduction

The emergence of Generative Artificial Intelligence (GenAI) has brought us to an exciting crossroads in the world of education, unlocking the potential for shared interactions between human expertise and artificially intelligent systems. With the advent of Large Language Models (LLMs) like ChatGPT, we've transitioned into a new era of lively and engaging learning experiences. Yet, our journey propels us further towards the new frontier of Large Multimodal Models (LMMs). These models, proficient in handling text, images, and audio, promise to deliver enriched, meaningful learning environments. The evolution envisages a landscape where AI blends seamlessly into our educational endeavors, potentially giving rise to proficient AI assistants that augment our digital interactions (Yang et al., 2023).

The education sector finds itself at a critical turning point, with growing uncertainty as learners will increasingly demand personalized, interactive learning experiences enabled by AI. The absence of a relevant framework for seamlessly integrating these emerging technologies into the broader educational processes and systems poses a challenge. This paper unfolds a unifying vision and charts a purposeful path forward to navigate the sense of escalating unease in the field of education, aiming to guide the integration of these emerging technologies into teaching and learning environments.

The Generative Education (GenEd) Framework, introduced in this paper, offers a practical path forward by blending traditional instructional methods with AI to create new types of AI-Blended Learning solutions and experiences. The GenEd Framework provides a structured approach for integrating AI in education, cultivating cooperative relationships between educators and AI, and guiding the necessary policy, infrastructure, and mindset changes to successfully navigate this technological transition. At its core, the GenEd Framework proposes AI-Learning Companions, transitioning AI from tools to collaborative partners in personalized learning. This framework sees every learning interaction, content, and material as valuable training data for AI models, combined with a method called Reinforcement Learning using Human Feedback (RLHF) which continuously improves the models' capabilities over time. In the GenEd Framework this feedback is education-centric, obtained from interactions among learners, educators, and AI. This specialized feedback, intricately tied to educational outcomes, enhances the AI models' learning efficacy.

Central to the GenEd Framework is establishing cooperative relationships between educators and AI. In this model, many educators will transition away from more traditional roles to become AI-Enhanced Mentors, providing insights that improve AI systems while still making vital contributions to the learners. This change envisions an educational setting where AI assists and augments, rather than replaces, human educators, creating a collaborative environment that significantly advances both access to and quality of education.

To drive this vision forward, we introduce Harmony, a practical product concept aimed at nurturing human connections in the AI-enhanced education world. Harmony represents an initial step towards realizing the GenEd Framework vision, demonstrating the harmonious potential of human and AI collaboration to create an effective and personalized learning environment.

This paper not only clarifies the GenEd Framework but also delves into the necessary changes in policy, infrastructure, and mindset needed to realize this vision. It unveils promising business opportunities ready to propel EdTech innovation and role redefinition within the education sector. Moreover, it extends the discussion to envision the significant impact and endless potential that future AI advancements, including LMMs, could have in reshaping the educational paradigm. In doing so, it underscores the need for proactive adaptation in the face of AI's ongoing evolution. Furthermore, this narrative explores the practical application and validation of the GenEd Framework's core principles, indicating a future of education that's not merely personalized but elevated by the constructive interplay of AI and human creativity.

# Background

## The Evolution of AI in Education Research

The path of blending artificial intelligence with education (AIEd) has seen a remarkable evolution. It began as early as the 1960s and 1970s with pioneering Intelligent Tutoring Systems (ITS) and Computer-Assisted Instruction systems. As we moved into the 1980s and 1990s, the field became more established with significant milestones like the launch of the International Journal of Artificial Intelligence in Education in 1989 and the creation of the International AI in Education Society (IAIED) in 1993. Over the last quarter-century, the focus has largely been on creating one-on-one tutoring systems, aiming to solve a challenge known as the "two-sigma problem," which is further discussed in the next subsection (Bloom, 1984).

Recent reflections on AI in education research from 1993 onwards have shed light on a wide range of focus areas. These range from instructional design and learning sciences to creating practical classroom tools and technological development. Technologies like Automated Assessment Systems, Adaptive Learning Platforms, and Multimodal Learning Analytics have played a crucial role in the past and continue to shape the educational landscape.

## Bridging the Two-Sigma Problem through AI-Blended Learning

The "Two-Sigma Problem," pinpointed by Benjamin Bloom, highlights a significant performance difference between students receiving one-on-one tutoring and those in traditional classroom settings. The GenEd Framework aims to bridge this gap through an AI-Blended Learning approach. By transitioning educators to AI-Enhanced Mentoring roles and introducing AI-Learning Companions, the framework seeks to provide personalized attention, similar to one-on-one tutoring, at a larger scale. This modern approach, fueled by advancements in AI like LLMs and LMMs, aims to address the Two-Sigma problem, thus aligning historical educational goals with modern AI-enabled possibilities.

## The Dawn of Large Multimodal Models in Education

The arrival of Large Language Models (LLMs) like ChatGPT initiated an exciting yet controversial chapter in the education sector, showcasing the profound capabilities of AI.

However, the exploration into AI-enhanced education is just beginning with LLMs as the initial step. The following wave promises to be so significant that it might fundamentally disrupt the entire idea and role of education in our lives. LLMs, mainly designed for text, fall short in capturing the full essence of human learning, which involves visual, auditory, and hands-on experiences.

Soon arriving are Large Multimodal Models (LMMs), a true game-changer not only for education but also set to cause dramatic changes across almost all industries. LMMs have the unique ability to process diverse types of data like text, images, videos, and speech all at once, creating a unified understanding (Yang et al., 2023). This approach will potentially allow LMMs to thoroughly understand a wide range of inputs, providing responsive and personalized learning support.

What really sets LMMs apart is their potential expertise in computer vision, which extends beyond mere object recognition to a deeper understanding of human emotions and interactions. They can not only understand the digital world but also glimpse into our physical world through video inputs, potentially reading a learner's emotional state, offering reassurance and support during challenging tasks. This capability opens up a realm where AI can provide not only intellectual but emotional support, tailoring responses and guidance based on a learner's current emotional state and learning pace. This transformation opens up new dimensions of learning, marking a significant leap in the educational landscape.

The ability of LMMs to understand our physical spaces through computer vision fits perfectly with spatial computing, shown by platforms like Apple's Vision Pro, opening new doors in education. Spatial computing blends the physical and digital worlds, turning any place into an intelligent classroom enhanced by AI instruction. The pairing of LMMs' computer vision with spatial computing's ability to interact with the physical world is set to fundamentally reshape education, merging the digital and physical realms in a unified learning environment.

For example, imagine an art history class studying Rembrandt where students enter a vivid 3D simulation of his world enabled by LMMs and spatial computing. The LMM tailors each student's perspective - a budding young artist may see Rembrandt incorporating their original artwork into a painting while discussing techniques and giving feedback. As the student explores a virtual recreation of Rembrandt's studio, they gain first-hand advice from the master artist on how to improve their skills. This mentoring from a personalized Rembrandt avatar, made possible by the LMM, provides invaluable insight that ignites the student's passion for painting. Immersing students in a tailored Rembrandt narrative promotes deeper connections with his art and life. This demonstrates how LMMs and spatial computing could redefine art education by constructing personalized simulated worlds that boost engagement, understanding, and inspiration.

This shift calls for strategic adaptation, and the GenEd Framework provides a vital roadmap. GenEd emphasizes reskilling educators, empowering them to provide diverse, high-quality training data to unlock the full potential of LMMs as personalized learning companions. This collaborative partnership between educators and LMMs aims to elevate education to new heights. By proactively embracing this transformation, education can harness the transformative capabilities of LMMs and spatial computing, catalyzing a future where learning breaks physical boundaries and is continuously enriched through human-AI collaboration.

## The Generative Education (GenEd) Framework

The emergence of advanced AI technologies like Large Language Models (LLMs) and Large Multimodal Models (LMMs) brings a transformative moment in education. The GenEd Framework lays out a practical roadmap to harness AI's potential to enhance, not replace, the contributions of education professionals. The ultimate goal is to create a collaborative environment involving not just educators and AI companions, but also institutions, governments, industries, and more, all working in harmony to enrich the learning experience. This collective endeavor aims to redefine education, positioning it at the crossroads of human expertise and cutting-edge AI technology. (see figure below)

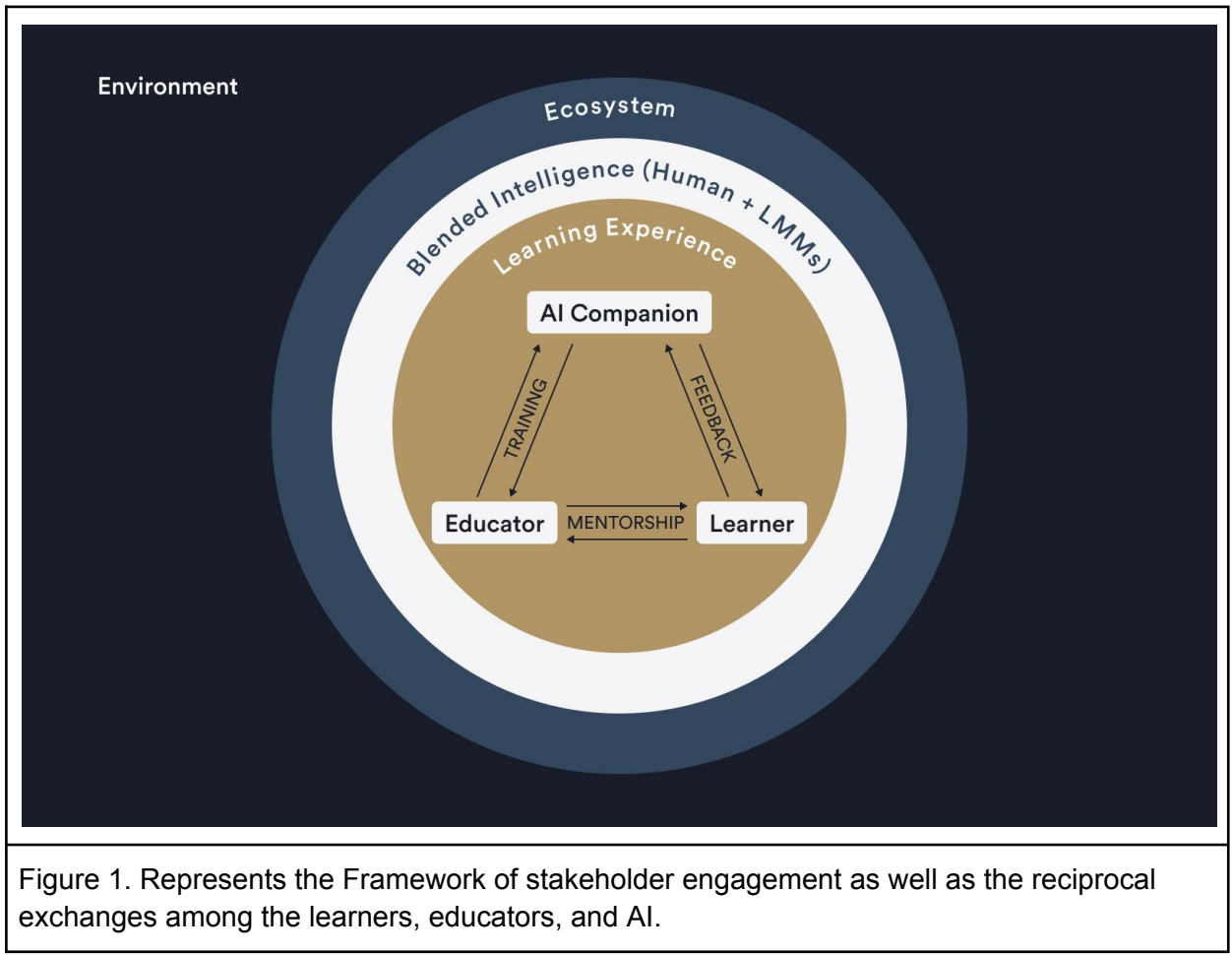

Figure 1. Represents the Framework of stakeholder engagement as well as the reciprocal exchanges among the learners, educators, and AI.

## Core Tenets

**Educator Focus Transition:**
Educators' roles shift from direct instruction to AI-Enhanced Mentoring, providing valuable inputs to expand AI capabilities and ensure AI companions effectively support learners.

**AI-Learning Companions:**
AI systems, trained by educators, serve as personalized learning companions, offering tailored support based on each learner's unique needs and pace.

**Reinforcement Learning with Human Feedback (RLHF):**
In the GenEd Framework this feedback is education-centric, obtained from interactions among learners, educators, and AI. This specialized feedback, intricately tied to educational outcomes, enhances the AI models' learning efficacy.

**Continuous Feedback Loops:**
Analyzing interactions between AI and students helps educators refine their inputs, ensuring AI companions align better with learning objectives and students' diverse needs (Chen & Zhu, 2022).

**Transformation of Education Systems:**
Broad changes across policy, infrastructure, and mindsets are crucial for the successful transition towards the GenEd Framework, including creating a supportive environment for educator reskilling, infrastructure development, and policy adaptations (Williamson & Hogan, 2021).

## Essential Contributions from Educators

The GenEd Framework emphasizes the critical role of educators in training robust AI systems. The contributions from educators form the foundation of a successful AI-Blended Learning environment, enabling AI to understand, interact, and adapt to varied learning needs. Below are the key categories of contributions along with a brief explanation of each:

| Educator Contribution Types | | | |
|---|---|---|---|
| **Category** | **Description** | **Components** | **Importance** |
| **Learning Content** | Structured and unstructured educational materials required for teaching. | Lessons, quizzes, assignments, etc. | Forms the core knowledge base for AI to support student learning. |
| **Learning Media** | Media resources utilized to enhance teaching and learning. | Videos, audio recordings, interactive simulations, infographics, etc. | Provides a diverse set of resources for AI to cater to different learning styles, enhancing engagement and retention. |
| **Practical Experiences** | Real-world examples and experiences enriching the learning context. | Case studies, field experiences, project-based learning, etc. | Enhances AI's ability to relate abstract concepts to real-world scenarios, aiding comprehension and retention. |
| **Professional Expertise** | Subject matter and pedagogical expertise of educators. | Subject matter knowledge, pedagogical techniques, classroom management strategies, etc. | Enriches AI's teaching capability with professional insight, ensuring effective instructional delivery. |
| **Learner Feedback** | Data reflecting student interactions, feedback, and engagement. | Student performance data, engagement metrics, feedback on materials, etc. | Enables AI to adapt instruction based on individual learning behaviors and preferences, ensuring personalized support. |
| **Creative Strategies** | Innovative teaching methods and creative problem-solving frameworks. | Innovative teaching methods, engaging learning activities, creative problem-solving frameworks, etc. | Drives AI's capability to introduce innovative, engaging, and effective learning strategies. |

## Implications for Educators and Institutions

Implementing the GenEd Framework requires significant shifts in education systems and mindsets, envisioning educators as empowered shapers of the AI-enhanced education landscape. (Holmes & Bialik, 2023). Key focus areas include:

<u>Integration Support:</u>

Providing educators the skills needed for AI-Enhanced Mentoring through online platforms, workshops, and cultivating continuous learning.

Incentive Structures:
Establishing programs recognizing educators' vital role in learner success enabled by AI and crafting incentives around improved student outcomes.

Process Adaptations:
Streamlining workflows to facilitate educator-AI collaboration time focused on elevating learning and teaching.

Technology Infrastructure:
Developing platforms that enable effective contribution delivery, and providing resources like high-speed internet and sufficient computing to support AI integration.

Policy and Funding:
Evolving policy frameworks to support AI-Blended Learning, promoting funding for R&D and pilot programs to validate optimal human-AI collaboration in education..

Realizing the Potential:
Developing strategies to engage an expanding learner base, creating collaborative platforms for interaction and co-learning among educators, learners, and AI systems.

Monitoring and Evaluation:
Establishing metrics to evaluate the effectiveness of AI-enhanced education initiatives, implementing feedback mechanisms for continuous improvement.

Community Engagement:
Involving all stakeholders in the transition towards AI-Blended Learning. Conducting public awareness campaigns about the benefits and challenges of AI-Blended Learning.

By addressing these areas, educational institutions can create a conducive environment for implementing the GenEd Framework, ensuring both educators and learners are well-prepared to harness the benefits of AI in education.

## Challenges and Solutions

Educators might face hurdles like lack of technical expertise or resources. Solutions could include professional development programs, collaborative platforms for resource sharing, and institutional support in terms of time and technology (Holmes & Bialik, 2023).

**Tools and Resources:**
Various tools and platforms can assist educators in making these contributions, like collaborative platforms for resource sharing, AI training platforms, and digital repositories for educational materials.

**Measurement and Evaluation:**
The effectiveness of contributions can be evaluated through the performance of the AI, student engagement metrics, and feedback from both students and educators.

**Continuous Improvement:**
Feedback on AI performance and student outcomes can refine and enhance future contributions from educators, creating a loop of continuous improvement.

**Ethical Considerations:**
Ethical considerations like data privacy, consent, and bias mitigation are paramount when training AI systems (Kaktiņš, 2023).

**Technological Support:**
Adequate technological infrastructure and support are crucial, including platforms for AI training, data analytics tools, and secure data storage solutions (Liu & Li, 2023).

## Product Concept: Harmony

**Crafting Personalized Learning Journeys with a Compassionate AI Companion**

Harmony, born from the Generative Education (GenEd) Framework, presents itself as a nurturing companion in the realm of education. This concept carves out a warm space where human educators and Harmony, the attentive AI companion, come together to elevate learning experiences. By morphing the gentle ideas of GenEd into a practical blueprint, Harmony sets the stage for rich and rewarding learning challenges.

**The Tale of Two Journeys**

In London, young Alex, an aspiring astronaut, starts his day facing algebraic hurdles. In Seoul, Ms. Park, a devoted educator, prepares for an afternoon of academic guidance. Their worlds converge through the Harmony framework, a digital realm where learning dances hand in hand with imagination.

In her serene workspace in Seoul, Ms. Park logs into her interactive dashboard, a gateway to her students' academic quests. A notification from Harmony, acting almost as a thoughtful colleague, nudges her attention towards Alex, who seems to be grappling with algebraic equations. Harmony has been attentively monitoring Alex's progress and detects a stumbling block in his understanding of algebra, suggesting a touch of human intervention might clear the mist.

With a click, Ms. Park enters Alex's digital learning space. Sky, the AI companion Alex affectionately named for its pilot's attire reflecting the boundless nature of education, recognizes Ms. Park and seamlessly integrates her into the interactive session. Together, they dissect the algebraic problem, simplifying it for Alex. It's a blend of Ms. Park's seasoned expertise, Sky's personalized approach, and Alex's enthusiastic engagement that clears the path to understanding.

The Harmony framework goes beyond personalizing learning experiences. It promotes a global learning community by facilitating interactions between students from different parts of the world. Under Ms. Park's guidance, Alex collaborates with other students, engaging in group projects and discussions, expanding his horizons and learning from diverse perspectives.

Harmony also prioritizes tasks for Ms. Park based on her preferences and the immediate needs of her students, enabling her to focus more on engaging with students rather than administrative tasks. Ms. Park's contributions, in turn, help train Harmony's AI, enhancing its ability to assist in the learning process.

A significant aspect of Harmony is its utilization of Large Multimodal Models (LMMs). These LMMs process the vast amount of educational data, adapting to the learning patterns of each student and teacher, and providing personalized insights and recommendations. They interpret Ms. Park's teaching methods and adapt the platform to better support her and her students.

The delightful smile on Alex's face as comprehension dawns, symbolizes more than a personal achievement. It's a testament to the beautiful harmony of human and AI efforts under the gentle oversight of Harmony, opening doors to a collaborative and enriching learning experience.

**A Glimpse into the Future**

Harmony opens a door to a place where the collaborative spirit of the GenEd Framework materializes into a tangible reality. This narrative unveils a snippet of the potential within Harmony's concept, hinting at a union of technical sophistication and learner-centric values. As we delve deeper into this space, we encounter new terminologies like Linguistic Learner Interfaces (LLIs) - a shift towards more natural interactions with learning platforms, and Perpetual Learning Continuity (PLC) - a strategy for seamless learning across different environments and devices. These newly coined concepts, LLIs and PLC, will be explored in depth alongside the technical product model and architecture in a technical white paper available in a separate publication.

## Challenges for Adoption

The bold dream of blending human and AI interactions in education, as outlined by the Generative Education (GenEd) Framework, unveils a journey filled with both challenges and promises. Moving towards this new realm of AI-Blended Learning calls for navigating a tricky landscape of technical, ethical, and policy hurdles. This section outlines the key challenges and explores potential routes to create a nurturing and engaging educational journey, with the rise of Large Multimodal Models (LMMs) as a game-changing force.

**<u>Understanding and Preparation:</u>**
The details of actively contributing to AI training and interacting with AI companions require a basic understanding and technical skills among education professionals. Bridging this knowledge gap is crucial to fully embrace AI-Blended Learning (An et al., 2022; Alimisis, 2021).

**<u>Identity and Role Redefinition:</u>**

The rise of AI as attentive learning buddies reshapes the traditional roles of educators. Welcoming this shift towards AI-Enhanced Mentoring roles is essential for fostering a harmonious relationship between humans and AI in education (Zhu et al., 2023).

**Workflow and System Adaptations:**
Integrating AI into educational workflows calls for rethinking existing systems and processes. Tailoring workflows for smooth human-AI collaboration and continuous improvement is key to realizing the envisioned GenEd Framework (Zhou et al., 2020).

**Technical Skill Acquisition**
Providing education professionals with the necessary technical skills for meaningful AI interactions creates a conducive environment for AI-Blended Learning (Zhang et al., 2023).

**Incentive Structures:**
Creating incentive structures that acknowledge and reward the efforts of education professionals in AI training and collaboration is vital for nurturing a culture of continuous improvement and engagement.

**Privacy, Ethics, and Consent**
Tackling the ethical aspects of AI in education, ensuring data privacy, and obtaining informed consent are critical for building trust and ensuring responsible AI deployment (Shah et al., 2023).

**Infrastructure and Resource Availability**
Sufficient investment in infrastructure and resources is crucial for smooth AI integration and effective implementation of the GenEd Framework (Chen et al., 2021).

**Policy and Regulatory Environment**
Adapting policy frameworks to support AI-Blended Learning, ensuring alignment with educational standards, and encouraging cross-border collaborations are critical for creating a supportive regulatory environment (Zawacki-Richter et al., 2023).

**Ethical Assessment and Evaluation**
Addressing ethical considerations in AI-driven assessments and evaluations, promoting fairness, and ensuring transparency are key to maintaining accountability and trust in AI-Enhanced Education (Nguyen et al., 2023).

**Cross-Sector Collaboration**
Involving governmental, educational, and private sector bodies in collaborative policy creation, ethical discussions, and technological innovations is essential for holistic development and successful adoption of AI-Blended Learning (Tochner et al., 2020).

Collective efforts from educational institutions, policymakers, and the AIED community are vital to overcome these challenges and pave the way for realizing the immense potential of human-AI collaboration in education, including the transformative capabilities offered by Large Multimodal Models (LMMs).

# Practical Applications

The Generative Education (GenEd) Framework unveils a spectrum of possibilities. Now we will delve into additional practical aspects of this framework, as it's vital to anchor our discussion in real-world examples, concrete implementations, and the potential ripple effects across the educational scene.

This segment, "Practical Applications," ventures into the GenEd Framework from various angles: a case study that illustrates a step toward the envisioned friendly interaction and a glimpse into additional paths for industry innovations.

## Case Study: Knewton's Alta Platform

The journey towards embracing the principles of the Generative Education (GenEd) Framework shines through in existing edtech platforms. A standout example is Knewton's Alta platform, which gives us a taste of how personalized, AI-enhanced learning experiences can come to life.

Alta is a nimble learning platform that employs AI to shape the learning journey for each student, based on their unique strengths, weaknesses, and pace. It's a treasure chest of educational content, with real-time data analytics at its heart, fine-tuning the learning experience continually.

**Integration of GenEd Principles**

Educator Focus Transition: In Alta, educators morph from merely delivering content to becoming key players in curating and refining the educational material. Their expertise fuels a rich, varied repository that powers the AI algorithms, aligning with GenEd's focus on AI-Enhanced Mentoring.

AI as Learning Companions: The platform's AI isn't just a tool; it's a companion in the learning journey, offering personalized support and feedback, bringing to life GenEd's idea of AI-Blended Learning

Continuous Feedback Loops: With real-time analytics, educators unlock insights into student performance and engagement, forming a continuous feedback loop that helps polish both the AI system and their teaching strategies.

Transformation of Education Systems: Embracing the Alta platform invites a shift in mindset and infrastructure, resonating with GenEd's call for sweeping changes across policy, infrastructure, and attitudes.

This case study shines a light on a step toward the harmonious relationship between AI and educators as envisioned in the GenEd Framework. While Alta might not capture the full essence of GenEd principles, it showcases a practical step toward weaving AI into education to enrich personalized learning and educator involvement. By dissecting the successes and roadblocks faced in Alta's implementation, stakeholders can harvest valuable insights into the practical steps and considerations crucial for marching towards the GenEd Framework.

## Additional Industry Innovations

Building on the principles showcased in Harmony, the GenEd Framework takes its innovative spirit into various corners of the education industry. It aims to nurture a cooperative space between AI and human engagement, creating a foundation for new business ventures and tech advancements that could make the education landscape even richer. Here's a glimpse into some promising innovations aligned with the collaborative spirit of the GenEd Framework:

<u>AI-Enhanced Content Creation Platforms:</u> With the rise of Large Language Models (LLMs), there's potential for platforms that help educators swiftly craft high-quality instructional materials, meeting a crucial need in adult learning, training, and upskilling without compromising accuracy or clarity (Leiker et al., 2023). By streamlining the content creation process and ensuring alignment with educational standards, these platforms can help educators focus more on personalized interaction with students (Zawacki-Richter et al., 2023; Kaktiņš, 2023).

<u>AI-Powered Assessment Solutions:</u> By merging AI with assessment tools, we can provide real-time, personalized feedback for students, creating a more dynamic and responsive learning environment while giving educators actionable insights.

<u>Personalized Learning Pathways:</u> The GenEd Framework encourages a setting conducive to personalized education. The goal is to create platforms that map out individualized learning pathways, with AI continuously adapting to each student's evolving needs and performance.

<u>Collaborative Learning Environments:</u> Digital spaces where students, educators, and AI interact in real-time could make learning more engaging and effective, possibly extending to realms like AI-facilitated group projects and peer-to-peer learning.

<u>AI-Driven Career Guidance:</u> Utilizing AI to analyze market trends, skills demands, and individual student capabilities for personalized career guidance could be a valuable service, helping students make informed decisions about their education and career paths.

<u>EdTech Consultancy and Integration Services:</u> As educational institutions journey towards the GenEd Framework, there's likely to be a demand for consultancy and integration services to help navigate the technological, policy, and workflow adaptations needed.

<u>Continuous Professional Development Platforms:</u> Platforms dedicated to ongoing professional development for educators, focusing on the changing narrative of AI in education, are pivotal to ensuring educators remain adept at using AI for better teaching and learning experiences (Zawacki-Richter et al., 2023).

<u>Ethics and Bias Mitigation Tools:</u> Developing tools and platforms to address potential biases in AI-driven education, promoting fairness, and ethical AI use represents a significant area for innovation.

The GenEd Framework lays the groundwork for these and many other innovations, each with the potential to contribute significantly to the evolving educational landscape. By embracing the collaborative spirit of the GenEd Framework, stakeholders across the education and technology sectors can work together to explore these innovative horizons, leading to a more enriched, accessible, and effective learning experience for all (Zawacki-Richter et al., 2023; Kaktiņš, 2023).

## Conclusion and Future Outlook

The meeting point of AI technologies with education, as envisioned in the Generative Education (GenEd) Framework, heralds a new era of collaborative and enriched learning experiences. In this evolving scene, educators are transitioning into AI-Enhanced Mentors, and AI becomes personalized Learning Companions, together enhancing the learning journey (Zawacki-Richter et al., 2023; Leiker et al., 2023).

Realizing GenEd's vision requires collective efforts across the educational ecosystem. It calls for strategic investments in reskilling educators, building robust technical infrastructure, and creating supportive policy frameworks (Zawacki-Richter et al., 2023). Initiatives like pilot programs and innovative projects such as Harmony are pivotal in validating and refining GenEd, paving the way for widespread adoption.

As AI continues to advance rapidly, staying ahead of the curve becomes crucial. The GenEd Framework serves as a guide for stakeholders, encouraging collaboration to harness AI's potential and elevate the human aspect of education, fostering enhanced learning experiences (Zawacki-Richter et al., 2023; Leiker et al., 2023).

The journey ahead is full of opportunities and challenges. It calls for a united effort from educators, technologists, policymakers, and all stakeholders to redefine education in the AI era. The GenEd Framework is more than an idea; it's a call to action. This call resonates with the aspiration to keep education centered on humanity, enhanced by the thoughtful integration of AI, nurturing a culture of continuous learning, growth, and discovery.

Recognizing the importance of emerging technologies like Large Multimodal Models (LMMs) and spatial computing is crucial as we move forward, as they can elevate educational experiences to new heights. Therefore, educational institutions, policymakers, and stakeholders must strategically adapt, prioritize educator reskilling, and fully harness the potential of LMMs as personalized learning companions. This path leads to a future where learning transcends traditional boundaries, becoming a continuous, adaptable, and enriched endeavor.

## Glossary of Terms

| Term | Definition |
|---|---|
| **Adaptive Learning** | A learning model that personalizes educational content and pathways based on the individual learner's performance, preferences, and pace. |
| **AI-Learning Companions** | The cooperative interaction between learners and AI systems, wherein AI serves as a supportive companion to facilitate personalized learning experiences. |
| **AI-Blended Learning** | An educational approach that combines traditional teaching methods with AI technologies to create a more personalized and interactive learning experience. |
| **AI-Enhanced Content Creation Platforms** | Platforms that leverage AI technologies to assist educators in creating instructional materials, streamlining the content creation process and ensuring alignment with educational standards. |
| **AI-Enhanced Mentoring** | The role of educators in providing personalized guidance and support to students while leveraging AI technologies to enhance the learning process. |
| **AI-Powered Assessment Solutions** | Tools and platforms that utilize AI to provide real-time, personalized feedback on learner performance, aiding in the continuous improvement of both teaching and learning processes. |
| **Artificial Intelligence (AI)** | The simulation of human intelligence processes by machines, especially computer systems, including learning, reasoning, and self-correction. |
| **Collaborative Learning Environments** | Digital or physical spaces designed to foster collaborative learning among students, educators, and sometimes AI, through interactive discussions, group projects, and peer-to-peer interactions. |
| **EdTech (Educational Technology)** | The use of technology, including software, hardware, and processes, to enhance educational experiences, improve outcomes, and streamline administrative tasks. |
| **Ethics and Bias Mitigation in AI** | Strategies and tools aimed at identifying, addressing, and reducing biases and ethical concerns in AI systems, ensuring fairness and promoting responsible AI use. |
| **Generative AI** | A type of artificial intelligence that can generate new data similar to the data it was trained on, often used in creating realistic text, images, or audio. |
| **Generative Content** | Content created or generated by AI systems, often based on data inputs, user interactions, or pre-defined criteria. |
| **Generative Education (GenEd) Framework** | A framework proposing the symbiotic collaboration between educators and AI, focusing on transitioning educators towards AI-Enhanced Mentoring roles to enhance learning experiences. |
| **Large Language Models (LLMs)** | AI models capable of understanding and generating human-like text based on the input they receive. |
| **Large Multimodal Models (LMMs)** | AI models capable of understanding and generating content across multiple modalities such as text, images, and audio. |
| **Learning Experience Platforms (LXP)** | Online platforms designed to provide personalized, continuous learning experiences, often encompassing content aggregation, AI-driven recommendations, and social learning features. |
| **Personalized Learning** | An educational approach tailored to the individual needs, preferences, and performance of each learner, often facilitated by AI and data analytics. |

| **Real-Time Analytics** | The process of analyzing and reporting on data as soon as it is collected, allowing for immediate insights and decision-making. |
|---|---|
| **Reinforcement Learning using Human Feedback (RLHF)** | A machine learning strategy where AI systems learn and improve through human feedback, aligning the AI's performance with human values and objectives. |
| **Spatial Computing** | The convergence of physical and digital worlds, using the space around us as a medium to interact with digital or virtual content. |